%% file: bias-sr.tex
\documentclass{sig-alternate-10pt-pods09}

\usepackage{subfig}
\usepackage{graphicx}

\usepackage{hyperref}

\newcommand{\varvspace} {\vspace{-0.16cm}}

\newtheorem{definition}{Definition}

\def\biasmeter{\sc{BiasMeter}}
\newcommand{\IP}[0]{OIP}
\newcommand{\IPs}[0]{OIPs}
\begin{document}

\title{On Measuring Bias in Online Information}
\numberofauthors{2}

\author{
	\alignauthor
	Evaggelia Pitoura, Panayiotis Tsaparas\\
	       \affaddr{Department of Computer Science and Engineering }\\
	       \affaddr{University of Ioannina }\\
	       \email{\{pitoura,tsap\}@cs.uoi.gr}
	\alignauthor
		  Giorgos Flouris, Irini Fundulaki, Panagiotis Papadakos \\
	  \affaddr{Institute of Computer Science, FORTH }\\
	       \email{\{fgeo,fundul,papadako\}@ics.forth.gr} 
	       \and
	       \alignauthor
	        Serge Abiteboul \\
	          \affaddr{INRIA \& ENS Cachan} \\
	          \email{serge.abiteboul@inria.gr} 
	       \alignauthor Gerhard Weikum \\
	        \affaddr{Max Planck Institute for Informatics, Germany}\\
	       \email{weikum@mpi-inf.mpg.de}
}
\toappear{}	

\maketitle

\begin{abstract}
Bias in online information has recently become a pressing issue, with search engines, social networks and recommendation services being accused of exhibiting some form of bias.
In this vision paper, we make the case for a systematic approach towards measuring bias. To this end, we discuss formal measures for quantifying the various types of bias, we outline the system components necessary for realizing them, and we highlight the related research challenges and open problems. 
\end{abstract}

\input{introduction}

\input{related}

\input{types}

\input{metrics}

\input{components}

\input{challenges}

\input{conclusions}

\bibliographystyle{abbrv}

\varvspace
{\fontsize{8.5}{9.5}\selectfont
\bibliography{bias}}

\end{document}

%% file: introduction.tex
\section{Introduction}

We live in an information age where the majority of our diverse
information needs are satisfied online by search engines, social
networks and media, news aggregators, e-shops, vertical portals, and
other online information providers (\textbf{\IPs}). For every request we submit to
these providers, a combination of sophisticated
algorithms produce a ranked list of the most relevant results tailored
to our profile.
These results
play an important role in guiding our decisions (e.g., where should I dine, what should I buy, which jobs should I apply to), in shaping our opinions (e.g., who should I vote for), and in general in our view of the world.

Undoubtedly, the various {\IP}s help us in managing and exploiting the abundance of available information.
But, at the same time, the convenient and effective way in which the {\IP}s  satisfy our information needs has limited our information seeking abilities, and has rendered us overly dependent on them. We rarely wonder whether the returned results represent all different viewpoints, and we seldom escape the echo chambers and filter bubbles created by personalization. We have come to accept such results as the ``de facto'' truth.

There are increasingly frequent reports of {\IP}s exhibiting some form of bias. For instance, in the recent US presidential elections, Google was accused of being biased against Donald Trump\footnote{https://www.theguardian.com/us-news/2016/sep/29/donald-trump-attacks-biased-lester-holt-and-accuses-google-of-conspiracy} and Facebook of contributing to the post-truth politics\footnote{https://www.theguardian.com/us-news/2016/nov/16/facebook-bias-bubble-us-election-conservative-liberal-news-feed}. Google search has been accused of being sexist or racist when returning images for queries such as ``nurse" or ``hair-styling"\footnote{http://fusion.net/story/117604/looking-for-ceo-doctor-cop-in-google-image-search-delivers-crazy-sexist-results/}, and prejudiced when answering queries about holocaust\footnote{http://www.bbc.com/news/technology-38379453}. Similar accusations have been made for Flickr, 
Airbnb and LinkedIn. 
In fact, the problem of understanding and addressing bias 
is considered a high-priority problem for machine learning algorithms and AI for the next few years\footnote{https://futureoflife.org/ai-principles/}.

The problem has attracted some attention in the data management community as well \cite{stoyanovich2016data}.
In this paper, we make the case for a systematic approach to addressing the problem of bias in the data provided by the {\IP}s. Addressing bias involves many steps. Here, we focus on the very first step, that, of defining and measuring bias.

According to the Oxford English Dictionary\footnote{https://en.oxforddictionaries.com/definition/bias.}, bias is 
\textit{``an inclination or prejudice for or against one person or group, especially in a way considered to be unfair''}, and as  \textit{``a concentration on or interest in one particular area or subject''}.
When it comes to bias in  \IPs, we make the distinction between  \textit{user bias} and \textit{content bias}. 
User bias appears when different users receive different content based on  user attributes that should be protected, such as gender, race, ethnicity, or religion. Content bias refers to biases in the  information received by any user, such as, when some aspect is disproportionately represented in a query result or in news feeds.

In the rest of this paper, we present the related work, describe formal measures for both user and content bias, and outline the basic components of a system for realizing these measures. Finally, we provide a synopsis of  challenges in identifying bias in online information.

%% file: related.tex
\section{Related Work}

In the field of machine learning, there is an increasing concern about the potential risks of data-driven approaches in  
decision making algorithms \cite{barocas2014datas,BolukbasiCZSK16a,hajian2016algorithmic,lepri2016tyranny,romei2014multidisciplinary,stoyanovich2016data}, raising a call for equal opportunities by design \cite{house2016big}. 
Biases can be introduced at different stages of the design, implementation, training and deployment of machine learning algorithms.
There are reports for  discriminatory ads based on either race \cite{skeem2016risk,sweeney2013discrimination}, or gender \cite{fairness-awareness}, and recommendation algorithms showing different prices to different users \cite{hannak2014measuring}. AdFisher \cite{datta2015automated} runs browser-based experiments to explore how user behaviors and profiles affect ads and if they can lead to seemingly discriminatory ads. Consequently, there are efforts for defining principles of accountable algorithms\footnote{http://www.fatml.org/resources/principles-for-accountable-algorithms}, for auditing algorithms by detecting discrimination \cite{sandvig2014auditing} and for debiasing approaches \cite{de-biasing, DBLP:journals/corr/ZhaoWYOC17}. There is
a special interest for racial fairness and fair classifiers \cite{hardt2016equality,zafar2015learning,zafar2017www-fairness,corbett-DaviesP17},  ensuring that groups receive ads based on population proportions \cite{fairness-awareness} and reducing the discrimination degree of algorithms against individuals of a protected group \cite{fish2016confidence}. Other efforts try to ensure temporal transparency for policy changing events in decision making systems  \cite{ferreira2016case}. Recently, tools that remove discriminating information\footnote{http://www.debiasyourself.org/}, help in understanding opposing opinions\footnote{https://www.escapeyourbubble.com/}, flag fake news\footnote{http://www.theverge.com/2016/12/15/13960062/facebook-fact-check-partnerships-fake-news}, increase transparency of personalization algorithms\footnote{https://facebook.tracking.exposed/}, or show political biases of Facebook friends and news feed\footnote{http://politecho.org/} have started to appear.

Another branch of research focuses on how bias can affect users. According to field studies, users of search engines trust more the top-ranked search results \cite{pan2007google} and biased search algorithms could shift the voting preferences of undecided voters by as mush as 20\% \cite{seme-elections}. Since most users try to access information that they agree with \cite{koutra2015events}, the personalization and filtering algorithms used by search engines lead to echo chambers and filter bubbles that reinforce bias \cite{Bozdag13filtering,hannak2013measuring}. This is also evident in social media where platforms strengthen users’ existing biases \cite{liu2014twitter}, minimizing the exposure to different opinions  \cite{weber2013secular}. Rating bubbles emerge especially when positive social influence accumulates, while crowd correction neutralizes negative influence \cite{muchnik2013social}.

Previous studies have looked at individual aspects of bias, such as geographical (i.e. whether sites from certain countries are covered more) \cite{vaughan2004search}, or temporal  (recommending recent and breaking news) \cite{chakraborty2015can}. Other approaches try to examine how bias can be measured \cite{mowshowitz2005measuring} and if search engines can partially mitigate the rich-get-richer nature of the Web and give new sites an increased chance of being discovered \cite{fortunato2006topical}. The presence of bias in media sources has been studied based on human annotations \cite{budak2016fair} and by exploiting affiliations \cite{wong2013quantifying} and the impartiality of messages \cite{zafar2016message}, while \cite{kulshrestha2017quantifying} tries to quantify bias in Twitter data.
There is clearly a need for a systematic approach to identifying  bias in online information, and in this paper, we outline some required steps and related challenges to this end.

%% file: types.tex
\section{Types of bias}
We consider  bias in terms of \textit{topics}. 
In particular, we would like to test whether an \IP\ is biased with respect to a given topic.  
A topic may be a very general one, such as, politics, or a very specific one down to the granularity of a single  search query.
For example, we may want to test whether an \IP\ provides biased results for events such as``Brexit'' and ``US Elections'', people such as ``Donald Trump'', general issues such as ``abortion'' and ``gun control'',  transactional queries such as ``air tickets'', ``best burger'', or even topics such as ``famous people''. An \IP\ may be biased with respect to one topic and unbiased with respect to another one.

We distinguish between two types of bias, namely \textit{user} and \textit{content}  bias.
User bias refers to bias against the users receiving the information, while content bias looks at bias in the information delivered to users.

For user bias, we assume that some of the attributes that characterize the user of an {\IP} are \textit{protected} (e.g. race, gender, etc.).
User bias exists when the values of these attributes influence the results presented to users. For example consider the case of a query about jobs,
where women receive results of lowered paid jobs than men.
User bias can also appear due to hidden dependencies between protected and unprotected attributes, even when such protected attributes are
not  used directly in computing the results (e.g., see \cite{disparate-impact}). For instance, the home location of users may imply their race.

Content bias refers to bias in the results provided by the \IP\ and may appear even when we have just a single user. For example, an instance of this kind of bias occurs when an \IP\ promotes its own services
over the competitive ones, or, when the results for queries about a political figure take an unjustifiable favorable, or unfavorable position towards this politician (independently of the user receiving the results).

In most cases, the \IP\ content is presented in the form of a ranked list of  results.
Results are often complex objects, such as news feeds, web pages, or, even physical objects, in the case of recommendations. We assume that results can be described by features, or attributes, either explicitly provided, or intentionally extracted.
In analogy to protected attributes for users, we consider \textit{differentiating attributes} for topics. 
For instance, for a controversial topic such as ``abortion'' or ``gun control'', the differentiating attribute could be the stance (pro, or against).
For a topic such as ``famous people'', we may want to test whether the results are biased towards men over women, or, favor people from specific countries, or, over-represent, say, artists over scientists.
Finally, for a topic such as ``US Elections'', a differentiating attribute may be the political party (with values, ``Democrats'' or ``Republicans'').

In a sense, addressing user bias can be regarded as a counterweight to machine-learning and personalization algorithms that try to differentiate the needs of various user groups, so that these algorithms do not discriminate over specific protected attributes. On the other hand, addressing content bias has some similarity to result diversification \cite{DP10}. However, diversity is related to coverage, since we want all various aspects of a topic, even the rarest ones, to appear in the result. For content bias, we want the differentiating attributes to be represented proportionally to a specific ``ground truth''.

A commonly encountered case is the case of a combined user and content bias appearing when
a specific facet is over-represented in the results presented to a specific user population, e.g., democrats get to see more pro-Clinton articles than republicans.
This type of bias is also related to \textit{echo chambers}, i.e., the situation in which information, ideas, or beliefs are amplified, exaggerated or reinforced inside groups of equally-minded people. Since similar people may be interested in specific aspects of a topic, as a result the content they create, consume, or prefer is biased towards these aspects. Then, the information presented to them may reflect this bias and by doing so possibly amplify the bias, creating a bias-reinforcement cycle. In such cases, there is often some relation between the protected attributes of the users and the differentiating attributes of the topic.

%% file: metrics.tex
\section{Bias Measures}
\label{sec:metrics}

In this section, we present  measures for user and content bias. Our goal is not to be overly formal, but instead we provide such measures as a means to make the related research challenges  more concrete. 

We assume that the information provided by an \IP\ is in the form of a ranked list $R$.
In the core of each bias measure lies a definition of similarity between lists of results. 
For now, let us assume that given two ranked lists of results $R_1$ and $R_2$, there is a distance function $D_R(R_{1}, R_{2})$ that measures the distance between these two rankings. $D_R$ can be defined by employing existing distance metrics between ranked lists, or using a geometric embedding of the ranked lists that takes into account both the similarity between results in the list and the importance of their position. 
We will revisit this issue when we talk about content bias.

To simplify the discussion, in the following, we assume that the topic for which we 
want to measure bias is a single query $q$. We can generalize the definitions to a set of queries by adopting some aggregation measure of the metrics for a single query.

\vspace*{0.1in}
\noindent \textbf{User Bias.} 
Let $U$ be the \IP\ user population. 
For simplicity, assume a binary protected attribute that divides users into a protected class  $P$ and an unprotected class $\bar{P}$.  
For example, if the protected attribute is gender, $P$ may denote the set of women and $\bar{P}$ the set of men.
Intuitively, we do not want the information provided to users to be influenced by their protected attributes. 

The problem of user bias is somehow related to fairness in classification,
where individuals are classified in a positive or negative class.
Example applications include among others hiring, school admission, crime risk factor estimation,
medicine (e.g., suitability for receiving a medical treatment) and advertisement selection.

There are two general approaches to defining fairness, namely \textit{group} and \textit{individual fairness} \cite{fairness-awareness}.
Group fairness imposes requirements on the protected and unprotected class as a whole.
A common example of group fairness is  \textit{statistical parity}, where the proportion of members in the protected class that receive positive classification is identical to the proportion in the general population.
Individual fairness requires that similar people are treated similarly.
A problem with group fairness is that it does not take into account the individual merits of each group member and may  lead in selecting the less qualified members of a group. On the other hand, individual fairness assumes a similarity metric of the individuals for the classification task at hand. Such  metrics  are very hard to define.

A technical difference between fairness and user bias is that most work in fairness focuses on classification tasks, while, in our case, results are ranked.
Very recent work addresses fair ranking (where the output is a ranked list of individuals) by adopting a group based approach that asks for
a proportional presence of individuals of the protected class in all prefixes of the
ranked list \cite{julia-ranking,cikm17-ranking}.
A conceptual difference between the two problems is that in the case of fairness, users  are the ones who are being classified (or ranked), whereas in user bias, the users are the ones who receive ranked information.

An individual-based approach to user bias assumes that it is possible to define  an appropriate distance measure $D_{u}$ between the users in $U$. 
The distance should capture when two users are considered similar for the topic under consideration.
For instance, if the topic is jobs, individuals with the same qualifications should be considered similar independently of their gender. The following definition is based on the premise that similar users should receive similar result lists.

\begin{definition} [Individual User Bias]
	\label{def:individual}
	An online information 
	provider is individual user unbiased if for any pair of users $u_1$ and $u_2$, it holds:
	\begin{equation*}
		D_R(R_{u_1}, R_{u_2}) \leq 	D_u(u_1, u_2)
	\end{equation*}
where $R_{u_1}$ and $R_{u_2}$ are the result lists received by $u_1$ and $u_2$ respectively.
\end{definition}

There are many ways of capturing group-based user bias.
We will discuss one.
Let $\cal{R}_P$ be the union of the result lists seen by  the members of the protected class and
 $\cal{R}_{\bar{P}}$ be the union of the result lists seen by the members of the non-protected class.
We could aggregate the results in each of them to create two representative ranked lists, 
${R}_P$ and $R_{\bar{P}}$, for $\cal{R}_P$ and  $\cal{R}_{\bar{P}}$, respectively.
We can define user bias using these representative ranked lists.

\begin{definition} [Group  User Bias]
\label{def:group}
	An online information 
provider is group user unbiased if it holds:
\begin{equation*}
|D_R(R_{P}, R_{\bar{P}})| \leq \epsilon
\end{equation*}
for some small $\epsilon$ $\geq$ 0.
\end{definition}

Aggregating result lists is just one possibility. For instance, another definition is to require the probability that a member of $P$ receives any of the lists in $\cal{R}_P$ to be the same with the
probability that a member of $\bar{P}$ receives it (and vice versa). 
All group-based definitions ignore the profiles of individual users; i.e.,
they do not capture the fact that a result list should be relevant to the specific individual in the group who receives it.

\vspace*{0.1in}
\noindent \textbf{Content Bias.} 
Let us first assume that there is just one user.
Let $A$ be a  differentiating attribute, and
let $\{a_1,...,a_m\}$ be the values of $A$. For example, in the case of a query about elections, $a_1,...,a_m$ may correspond to the different parties that participate in the elections. We also assume that each result is annotated with the values of attribute $A$, meaning that the result is about these values. 

A distinctive characteristic of content bias is that it can be defined only relatively to some  ``ground truth'', or ``norm''.
But what is the ``ground truth''?
One option  is to consider the actual data used by the \IP\ 
for computing the content delivered to users as the ground truth. 
For example, this is the approach taken in \cite{kulshrestha2017quantifying} that compares  the political bias of Twitter search  with the bias in all tweets that contain the search terms.
However,
user-generated content may  include biases inflicted by  the design and affordances of the \IP\ platform, or by behavioral norms emerging on each platform.
Bias can also be introduced by the \IP\ during the acquisition of data (e.g. during
crawling and indexing for a search engine).
See \cite{social-data} for a complete analysis of the different biases and pitfalls associated with social data.
There are also cases, where the actual data used by the \IP\ may not be available, as with search engines. 

Ideally, we would like to have an indisputably unbiased ranked list of results.
Such lists could be constructed using an aggregation of \IPs\ and other external sources such as knowledge bases, or domain experts.
Crowdsourcing could also be utilized in creating such lists.
In some cases, an estimation of the distribution of values of the differentiating attributes in the general population may be available.
For example, for the election query, we could use external sources, such as polls, to estimate the actual party popularity and user intention to vote.
One could also think of creating bias benchmarks consisting of reference sample topics and result lists similar to TPC benchmarks for evaluating database system performance, and TREC
tracks for evaluating relevance in information retrieval. 

Given the  ground truth as an  ``ideal unbiased ranking'' $R_T$, we could define content bias looking at its distance from the ground truth.

\vspace{-0.05in}
\begin{definition} [content bias]
	\label{def:object2}
	An online information 
	provider is content unbiased if it holds:
	\begin{equation*}
	|	D_R(R_u,  R_T)| \leq \epsilon
	\end{equation*}
for some small $\epsilon$ $\geq$ 0.
\end{definition}

One way of defining  $D_R$ is by looking at the distribution of the values 
of the differentiating attribute in an ideal ranking.
Assume that we have  the ``ground truth''  in the form of probabilities $Pr_T(a_i)$ for all the attribute values which captures the relative popularity of each  value (e.g., the support of a party as measured by polls). 
Let $Pr(u, a_i)$ be the probability that user $u$ receives a result annotated with value $a_i$ (e.g., one possible definition is this to be defined as the fraction of the top-$k$ results that are about $a_i$). The following equation could serve as a definition of  $D_R$.

\vspace{-0.08in}
	\begin{equation}
		\label{eq1}
	D_R(R_u,  R_T) = \max_{i}	|Pr(u,a_i) -  Pr_T(a_i)| 
	\end{equation}

\noindent \textbf{Combined User-Content Bias.} 
We can refine user bias, using content-aware distance definitions, such as the one in Equation (\ref{eq1}).
For example, in Definition \ref{def:individual}, we could use:
\begin{equation}
\label{eq2}
D_R(R_{u_1},  R_{u_2}) = \max_{i}	|Pr(u_1,a_i) -  Pr(u_2,a_i)|
\end{equation}

Equation (\ref{eq2}) looks at the relative bias of the content seen by two users. Although both users may  receive biased content with respect to  ground truth, there is no user bias as long as content is equally biased. 

To look for echo chambers, we need to test the content bias in
the result lists seen by different users.
For instance, adopting the representative list approach to user bias, we may look at the distance of ${R}_P$ and $R_{\bar{P}}$ from the ground truth to test, for example, whether
specific attribute values are over-represented in the results shown to a population group.

%% file: components.tex
\section{A System for Measuring Bias}
\label{sec:Components}

\begin{figure}[ht]
	\centering
		\resizebox{1\columnwidth}{!}{
			\includegraphics{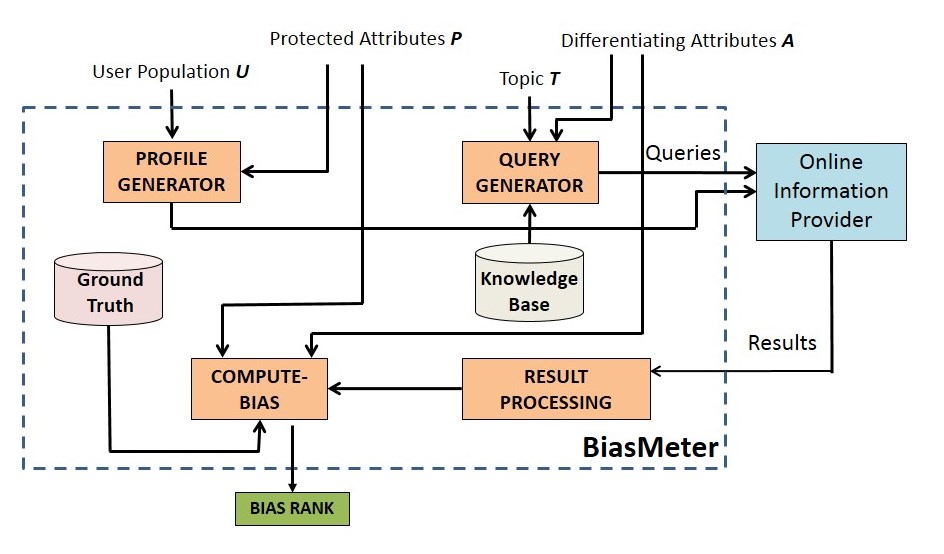}}
	\caption{System components.}
	\label{fig:components}
\end{figure}

We now look at some of the challenges involved in realizing a system for  measuring the bias of an \IP. The \IP\ may be a search engine, a recommendation service, the search or news feed service of a social network.
In Figure \ref{fig:components}, we present the main components needed by 
a system, called  {\biasmeter}, for measuring bias. 
We treat the {\IP} as a black-box and assume that {\biasmeter} can access it only through the interface provided by the {\IP}, e.g., through search queries.
For simplicity, we assume that the set of  protected  and 
differentiating attributes are given as input to {\biasmeter}.

Given the topic $T$ and the differentiating attributes  $A$, the goal of the	\textit{query generator} is to produce an appropriate set of queries to be submitted to the \IP\ under consideration. 
For instance, if the \IP\ is a search engine, to test about the topic ``US elections'',  the generator may produce a variety of search queries, including queries referring to specific political parties.
To produce queries that best represent the topic and the attributes, the query-generator may need to use background knowledge, such as, a related knowledge base. 
	
The \textit{profile generator} takes as input the user population $U$ and the set of protected attributes $P$ and produces as output a set of user profiles appropriate for testing whether the \IP\ discriminates over users in $U$ based on the protected attributes in $P$. 
For example, if we want to test gender bias in job search queries,
we need samples of men
and women, that have very similar characteristics with
respect to other attributes such as grades, skills, background, ethnicity, etc, to avoid differences that may appear due to attribute correlations.

There are many issues of both a theoretical and a practical nature in generating profiles. For example, we must ensure that the profiles are an appropriate sample of  $U$ that represents all values of the protected attributes. Furthermore, we should ensure that the characteristics of the users in the sample are similar with respect to all other attributes, so as to avoid the effect of confounding factors.  This raises issues similar to those met when selecting people for opinion polls, surveys, etc. 
From a more practical view, we need to assemble  users with the specific profiles and ask them to issue the queries (for example using a crowd-sourcing platform, such as Mechanical Turk), or generate artificial accounts of such users. 
An important step to automated profile generation is offered by AdFisher, a tool for testing discrimination in Google Ads \cite{datta2015automated}.
AdFisher builds user profiles by just using the Ad profile setting and by simulating visits at specific webpages. 

The \textit{result processing} component takes as input the results from the \IP\ and applies machine learning and data mining algorithms such as topic modeling and opinion mining to determine the values of the differentiating attributes. For example, if the topic is ``gun control'', we need to determine whether a specific result takes a positive, neutral or negative stand.

Finally, the \textit{compute-bias} component calculates the bias of the {\IP}, using our bias metrics and the \textit{ground-truth}.
Note that the cause of bias is
not specified in our result; we just detect bias with respect to specific user and content attributes.

%% file: challenges.tex
\section{Research challenges}
\label{sec:Challenges}

\noindent \textbf{Obtaining the ground truth.}
Defining the ground truth is the 
most formidable task in identifying bias.
One approach  could be a human-in-the-loop approach where users take the role of data processors 
characterizing the bias of online information, similarly to users evaluating the relevance
of search results.
One can even envision novel crowdsourcing platforms specifically targeting 
bias evaluation. 
However, such tasks are hindered 
by strong cognitive biases, such as confirmation bias, that may lead users in discrediting as biased  any information that does not fit their own believes. 
Furthermore, bias, as opposed to relevance, may involve political, ideological, or, even, ethical connotations. 
Besides crowdsourcing, one can envision a form of
data-driven validation that integrates information from large data repositories, knowledge bases, and  multiple \IPs.
Besides this long-term quest for ground truth, 
a more realistic approach is to rely on comparative evaluations.
For instance one could compare the bias between the results of two \IPs\, or between
the results of an \IP\ and content found in traditional media.

\noindent \textbf{Defining bias measures.}
Bias is multifaceted. We abstracted the many forms of bias, through the notions of protected attributes for users and differentiating attributes 
for content. 
However, there are often correlations among the attributes making it
hard to single out the effects of each of them in the results.
Further, our measures are high level, and a lot of work is needed to come up with
rigorous mathematical formulations.

\noindent \textbf{Engineering and technical challenges.} 
To measure bias with respect to a protected attribute $P$ (e.g. gender), we need to generate large samples of user accounts for the different values of $P$ (e.g., women and men), making sure that the distribution of the characteristics for the other attributes is near identical.  
Careful statistical analysis is also required to ensure statistical significance of our results.
In addition, the query generation and result processing components involve a variety of data mining and machine learning algorithms for identifying keywords to describe an information need, or understanding the topic and stance of a specific result. To this end, we need modules for knowledge representation, record linkage, entity detection and entity resolution, sentiment detection, topic modeling, and more.

\noindent \textbf{Auditing.} Bias detection can be  simplified, if access is given to the internals of the \IP\ (e.g., for sampling users with specific demographics, or getting  non personalized results). Clearly, this is impossible for an entity outside the \IP\ and it requires the cooperation of law and policy makers.
Such access would also help in differentiating between bias in the source data and bias in the results.
There is a growing literature advocating the systematic auditing of algorithms \cite{crawford2016can,pasquale2015black,sandvig2014auditing}.

%% file: conclusions.tex
\section{Conclusions}
In this paper, we argue about the importance of a systematic approach for measuring the bias of the information we get from online information providers. As more people rely on online sources to get informed and make decisions,
such an approach is of central value. Building  a system for measuring bias raises many research challenges, some of which we have highlighted in this paper.
Measuring bias is just the first step; many more steps are needed to counteract bias including  identifying bias sources and developing debias approaches.